# Experimental demonstration of the robust end state in a split-ring-resonator chain

Jun Jiang,[1,*] Zhiwei Guo,[1,*] Ya-qiong Ding,[1,2] Yong Sun,[1,†] Yunhui Li,[1] Haitao Jiang,[1] and Hong Chen[1,‡]

[1] *MOE Key Laboratory of Advanced Micro-structured Materials, School of Physics Science and Engineering, Tongji University, Shanghai 200092, China*

[2] *Science College, University of Shanghai for Science and Technology, Shanghai 200093, China*

**Abstract:** One of the fascinating topological phenomena is the end state in one dimensional system. In this work, the topological photonics in the dimer chains composed by the split ring resonators are revealed based on the Su-Schrieffer-Heeger model. The topologically protected photonic end state is observed directly with the *in situ* measurements of the local density of states in the topological nontrivial chain. Moreover, we experimentally demonstrate that the end state localized at both ends is robust against a varied of perturbations, such as loss and disorder. Our results not only provide a versatile platform to study the topological physics in photonics but also may have potential applications in the robust communication and power transfer.

---

[*] These authors contributed equally to the work
[†] yongsun@tongji.edu.cn
[‡] hongchen@tongji.edu.cn

# I. INTRODUCTION

The rise of topological photonics,[1,2] is along with the discovery of various topological phases in condensed matter physics. Topology, as a new concept in photonics, can be used to effectively control light-matter interaction and make use for robust one-way transmission. Recently, photonic topological one-way edge modes with broken time-reversal ($T$-reversal) symmetry have been widely studied in theory[3] and experiment[4-6] based on photonic crystals composed by gyro-magnetic materials and helical optical fibers, respectively. The quantum spin Hall effect (QSHE),[7,8] time-reversal symmetric electron systems with nontrivial topological properties, also attracted people's great attention for its novel spin-dependent topological phases. Inspiring by the QSHE of electron, up to now, a variety of optical analogue have been proposed, by use of metamaterials,[9-12] coupled ring resonators array,[13-15] synthetic gauge field,[16,17] two-dimension (2-D) photonic crystal.[18] In addition to the edge state in the 2-D system, the topological states in the one dimensional (1-D) system also have been demonstrated by metamaterials,[19] resonant dielectric structure,[20-22] photonic/phononic crystals,[23-25] and 1-D waveguide array.[26,27] The topological interface state between two crystals with distinct topological gap has been demonstrated, which can be used for the field enhancement.[25] Very recently, this interface state has been used to determine the topological invariants of the polaritonic quasicrystals,[28] and the chiral edge states in 1-D double coupled Peierls chain have been observed for the first time.[29]

Topological states are protected by the topological phase transition across the interface. Thus they are robust against the defects, the loss, and the disorder, which do not change the topological phase of the structure.[21,24,30] In 1-D system, the robustness of the topological interface state has been investigated in dielectric resonator chains and photonic crystals.[21,24] In this work, we experimentally demonstrate the robust end states of the Su-Schrieffer-Heeger (SSH) mode in a split-ring-resonator (SRR) chain. SRRs with broken rotational symmetry provide a new degree of freedom, the azimuth in addition to the distance and the background between the resonators, to adjust the coupling in the chain. Based on the SRR chains with limited length, we systemically studied the robustness of end states which are insensitive to a variety of perturbations, such as loss and disorder in the structure. Our results provide a versatile platform to observe the robust topological states in photonics. In addition, the robust end states at the two ends of a chain may have some potential applications in the information transmission, power transfer, topological gap soliton and so on.

The paper is organized as follows. In Sec. II, using the near-field method, we experimentally measure the density of states (DOS) for two dimer chains with different topological property, and demonstrate the end states exist in the topological nontrivial chain. After that, in Sec. III, by adding a variety of perturbations into the structure, we experimentally investigate the robustness of the end states in the topological nontrivial chain. Finally, we conclude in Sec. IV.

## II. PHOTONIC END STATES IN DIMER CHAINS COMPOSED OF SRRS

Our experimental setup is shown in the Fig.1. All the SRRs are identical with the same resonant frequency of $\omega_0 = 1.9$ GHz, which is determined by the geometric parameters, including the thickness $w = 1.0\,\text{mm}$, the height $h = 5.0\,\text{mm}$, the inner radius $r = 10\,\text{mm}$, and the gap size $g = 1.5\,\text{mm}$. The unit cell of the one-dimensional dimer chains consists of two SRRs whose gaps have different azimuth angles. The chain is composed of 16 dimerized unit cells in experiments. The sample is sandwiched by two metal plates in measurements (the top metal plate is removed to take a picture of the sample). The two metal plates separated by 30 cm act as a waveguide with the cutoff frequency of 5 GHz for polarization parallel to the plate. Below the cutoff frequency, there is only evanescence wave in bare waveguide. Therefore, the bare waveguide can be regarded as an electromagnetic "topological trivial insulator". In meanwhile, the coupling between SRRs can only rely on the near-field interaction due to the suppression of their far-field radiation. So our system works well in the tight-binding regime.

Considering only the nearest-neighbor coupling, the motion equation for the infinite dimer chain can be described as:

$$-\omega_{nor}^2 \begin{pmatrix} a_k \\ b_k \end{pmatrix} = \begin{pmatrix} -1 & \kappa_1 + \kappa_2 e^{-ikd} \\ \kappa_1 + \kappa_2 e^{ikd} & -1 \end{pmatrix} \begin{pmatrix} a_k \\ b_k \end{pmatrix}, \qquad (1)$$

Where $\omega_{nor}$ denotes the frequency, which is normalized with respect of the resonant frequency $\omega_0$, $\begin{pmatrix} a_k \\ b_k \end{pmatrix}$ represents the cell-periodic Bloch current Eigen-function of a state with wave-vector $k$ and lattice constant $d$. $\kappa_1$ and $\kappa_2$ are the normalized intra-dimer and the inter-dimer coupling strengths, respectively. Controllable coupling between resonators is the essential in experimental study of SSH model. A straightforward way is to resort to the dependence of coupling strength on distance.[21,31] Other methods include alternating the electromagnetic background between resonators [26] and utilizing the azimuthally dependent coupling between the dipoles [20,22]. In our setup, the rotational symmetry of the resonator is broken due to the gap, so that the coupling strength between SRRs in the chain can be flexibly adjusted by in-plane rotation of the SRRs only. Under a fixed separation, the coupling strength under the case where the gaps in neighboring SRRs are next to each other, is stronger than that where the gaps are on opposite sides.[32,33] When the distance between two SRRs is fixed to $d/2 = 24$ mm, the strong and the weak normalized coupling parameters in our setup are 0.48 and -0.21, respectively.

We first design two different dimer chains. For convenience, the chains with $|\kappa_1|>|\kappa_2|$ are called type I, and the ones with $|\kappa'_1|<|\kappa'_2|$ are called type II. They can emulate polyethylene ending with strong and weak bonds, respectively. We first consider the type I (topological trivial) chain with $\kappa_1=0.48$, $\kappa_2=-0.21$, whose unit cell is shown in the inset of Fig. 2(a). By using Eq. (1), the calculated Eigen frequencies of the type I chain composed of 16 unit cells is given with black dots in Fig. 2(a). One can find there are two isolated bands separated by a gap, which is indicated by the gray area. For this Type I sample, the measured DOS in the band is relatively high, whereas in the gap it is almost zero, as shown in Fig. 2(a). It is consistent with the theoretical calculation (marked by the black dots). Here the DOS spectrum is obtained by averaging the local density of states (LDOS) spectral over all sites, and the LDOS spectrum of each site is obtained from the reflection by putting the probe to the center of the corresponding SRR.[34] Here all of the LDOS measurements have been normalized.

Next, we studied a 16-unit type II chain (topological non-trivial) with $\kappa'_1 = -0.21, \kappa'_2 = 0.48$, whose unit cell is shown in the inset of Fig. 2(b). Similarly, we calculate its Eigen frequencies and measure its DOS spectrum, which consists with each other well in Fig. 2(b). Compared with the results of the type I chain, there is an additional state in the gap region for the type II chain. Calculations and measurements show that the LDOS of the new state is strongly localized at the two ends, as shown in Fig. 2(c). Hence it belongs to the end states. It is totally different from the ordinary state in the band whose LDOS is mainly distributed in the bulk [see in Fig.2 (d)].

The topological property of 1-D system can be characterized by the winding number of the band:[35]

$$w = \frac{1}{2\pi} \int_{\pi/d}^{-\pi/d} \theta_k dk, \qquad (2)$$

where $\theta_k$ is the polarization vector angle. After calculation, we get the winding number for both upper and lower bands are $w = 0$ for the type I chain, and $w' = 1$ for the type II chain (Details can be found in Appendix). According to the relationship between the band gap and the passband, the gaps of two chains considered above are completely different.[25] The band gap of type I chain is trivial as the bare waveguide, while the band gap of type II chain is nontrivial. The end states observed at the two ends of the type II chain are topologically protected by the topological transition - from trivial bare waveguide to nontrivial dimer chain. The robustness of the end states is investigated experimentally in the following section.

## III. EXPERIMENTAL DEMONSTRATION OF THE ROBUST END STATES AGAINST LOSS AND DISORDER

In this section, we will reveal that the end state in the type II chain is robust against certain loss and disorder. At first, we add the loss into the central 20 SRRs of the chain as shown in Fig. 3(a). The lossy SRRs marked by grey background are realized by adding absorbing materials into the interior of the rings. Measured DOS spectrum is shown in Fig. 3(b). One can clearly see that the ordinary bulk states are affected significantly while the end state in the gap region (grey area) is almost immune to the loss perturbations. In order to further illustrate the unique of the robust end state, the LDOS distributions in the lossless and loss chains are compared in Figs. 3(c) and 3(d). The measured LDOS of the end state is still confined at the two ends [the red triangles in Fig. 3(c)], just like that calculated in the lossless system [dashed line in Fig. 3(c)]. While for the bulk state, the LDOS is significantly affected, as shown in Fig. 3(d).

Secondly, we investigate the robustness of the end states against certain disorder perturbation. The structure disorder is realized by in-plane rotating the central 20 SRRs, as shown in Fig. 4(a). The detail of rotation is illustrated in the inset of Fig. 4(b). Three disorder chains are considered, in which the central 20 SRRs are random rotated of $\alpha = 1$, 3 and 5 degrees, respectively. Measured LDOS distributions of the end state and the bulk state in three chains are shown in Figs. 4(b) and 4(c), respectively. By comparing with the results of the original chain (dashed line), one can find that at different disorder levels the end state is still maintained, whereas the bulk state has been deteriorated seriously.

At last, the robustness of the end state is further examined by adding loss and structure disorder simultaneously. By comparing three structures with both loss and random rotate angles in the central 20 SRRs [just as the situation in Fig. 4(a)], we find the topological end state is still maintained, as shown in Fig. 5. Remarkably, our demonstration of topological robust effect using in-plane rotation of SRRs paves a new way to steer the random distribution, which does not resort to the structural disorder by randomly distributing the inter-site separations.[21]

In discussion, our results have revealed that the topological end state cannot be affected when the loss and the random rotation are introduced into the center 20 SRRs of the structure. This robust end state may have some significant applications, such as the robust communication and power transfer. When the electromagnetic wave is fed into one end of a chain (called end A), the end state will be established quickly. Then the electromagnetic field will accumulate at another end of this chain (called end B) and the electromagnetic field at end B can be interpreted as information or gathered as energy. In some special occasions, this robust topological end state will show great superiority to the cable transmission. For example, when the wire is buried deeply in the ground, the cable might be broken due to the geological activities or other natural disasters. The broken conductor cannot be restored, resulting in the termination of communication and power transfer. This potential risk can be greatly reduced by using the robust topological end state existing in the SSH chain by means of near-field coupling.

## IV. CONCLUSION

In summary, we experimentally demonstrate the end states of the SSH model by *in situ* measurements of the local density of states in a dimer SRR chain, in which SRRs with broken rotational symmetry provide the new azimuth degree of freedom to adjust the coupling between the resonators. It is observed the end states are robust against to a varied of perturbations, such as the loss and disorder. Our results not only provide a versatile platform to study the robust topological end state, but also may contribute potential application in the information transmission, power transfer, and so on. Although our results are from 1-D system, the array of split ring resonators even can be used to explore the topological phenomena in 2-D system.

## ACKNOWLEDGMENT

This research was supported by National Key Research Program of China (2016YFA0301101), National Natural Science Foundation of China (NSFC) (11674247, 61621001, 11504236, 11474220), Natural Science Foundation of Shanghai (No. 17ZR1443800), and the Fundamental Research Funds for the Central Universities.

**Appendix: Calculation of winding number**

For our system, the equation of motion in wave-vector space is governed by

$$\omega_{nor}^2 \begin{pmatrix} a_k \\ b_k \end{pmatrix} = D_k \begin{pmatrix} a_k \\ b_k \end{pmatrix}, \tag{S.1}$$

with $D_k = \begin{pmatrix} 1 & -\kappa_1 - \kappa_2 e^{-ikd} \\ -\kappa_1 - \kappa_2 e^{ikd} & 1 \end{pmatrix}$. The dynamical matrix $D_k$ play the same role of the Hamiltonian. The diagonal element is the onsite potential. $\kappa_1$ ($\kappa_2$) is dimensionless and denotes the intra-dimer (inter-dimer) coupling strength.

The eigenvectors of Eq. (S.1) for both upper and lower band are expressed as

$$|u_{k,\pm}\rangle = \begin{pmatrix} a_k \\ b_k \end{pmatrix} = \begin{pmatrix} \dfrac{\kappa_1 + \kappa_2 \cos(kd) - i\kappa_2 \sin(kd)}{\sqrt{(\kappa_1 + \kappa_2 \cos(kd))^2 + (\kappa_2 \sin(kd))^2}} \\ \mp 1 \end{pmatrix}. \tag{S.2}$$

Here we rewrite Eq. (S.2) as follows:

$$|u_{k,\pm}\rangle = \begin{pmatrix} e^{-i\theta_k} \\ \mp 1 \end{pmatrix}, \tag{S.3}$$

where $\theta_k = \arctan\left(\dfrac{\kappa_2 \sin(kd)}{\kappa_1 + \kappa_2 \cos(kd)}\right)$, and its normalized form is

$$|u_{k,\pm}\rangle = \frac{1}{\sqrt{2}} \begin{pmatrix} e^{-i\theta_k} \\ \mp 1 \end{pmatrix}. \tag{S.4}$$

The winding number is the number of loops made by a Bloch state around the equator of the Bloch sphere, as $k$ passes through the Brillouin zone. For our model, it can be expressed as follows:

$$w_{winding} = \frac{i}{\pi} \int_{-\pi/d}^{\pi/d} \left( a_k^* \partial_k a_k + b_k^* \partial_k b_k \right) dk. \tag{S.5}$$

Combine Eq. (S.4) and Eq. (S.5), we get

$$w_{winding} = \frac{i}{\pi} \int_{-\pi/d}^{\pi/d} \left( a_k^* \partial_k a_k + b_k^* \partial_k b_k \right) dk = \frac{1}{2\pi} \int_{-\pi/d}^{\pi/d} \theta_k dk \, . \qquad (S.6)$$

By numerical calculation, the winding numbers for both upper and lower band are 1 when $|\kappa_1| < |\kappa_2|$, while they are 0 when $|\kappa_1| > |\kappa_2|$.

**Figure Captions**

FIG. 1. Experimental setup. The one-dimensional dimer chain composed by equally spaced 32 identical SRRs (not shown all the SRRs) is arranged on a foam substrate, and sandwiched by two metallic plates in experiments (here the top metal plate is taken away in order to take a picture of the chain). The near-field probe made of a non-resonant loop is used to measure the density of states.

FIG. 2. Two types of dimer chains differ in their topological properties. (a-b) Calculated Eigen frequencies (black dots), and measured DOS spectrum (blue profile) of the Type Ⅰ and the Type Ⅱ chains are given in (a) and (b), respectively. The insets shows the corresponding schematics of the unit cells. (c-d) Experimental (red triangles) and theoretical (grey dashed line) LDOS distribution of the end state (c) and the bulk state (d), as indicated in (b).

FIG. 3. Robust end state against the loss perturbation. (a) The topological nontrivial chain with the loss added into the central 20 SRRs (indicated by the grey background). (b) Measured DOS spectrum with loss. DOS of the end state is much more robust than that of the bulk state. (c) Measured LDOS distribution of the end state with loss (red triangles), along with the theoretical calculations without loss (grey dashed line). (d) Measured LDOS distribution of the ordinary bulk state with loss (red triangles), along with the theoretical calculations without loss (grey dashed line).

FIG. 4. Robust end states against the disorder perturbation. (a) Schematic of the topological nontrivial chain with disorder (random coupling strengths by rotating the central 20 SRRs). (b, c) Measured LDOS distributions of the end state and the bulk state at various disorder levels, including $\alpha=1°$ (red circles), $\alpha=3°$ (blue stars), and $\alpha=5°$ (green triangles). Detail of rotation is shown in the inset of Fig. 4(b). As a comparison, the calculated LDOS distribution of the end and the bulk states in the original chain are also presented (gray dashed line).

FIG. 5. Measured LDOS distributions of the end state in the chain with both loss and disorder perturbations in the central 20 SRRs. As a comparison, the calculated LDOS distribution of the end state in the original chain are also presented (grey dashed line).

**Figures**

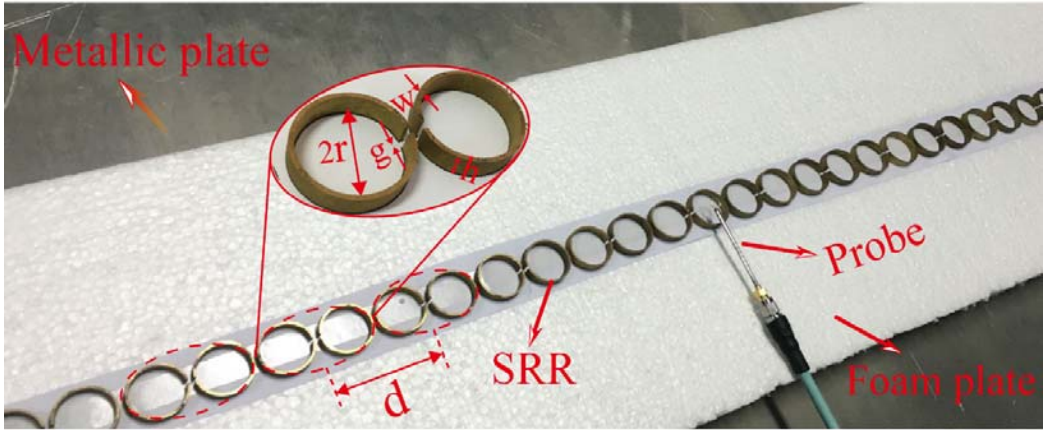

Fig. 1

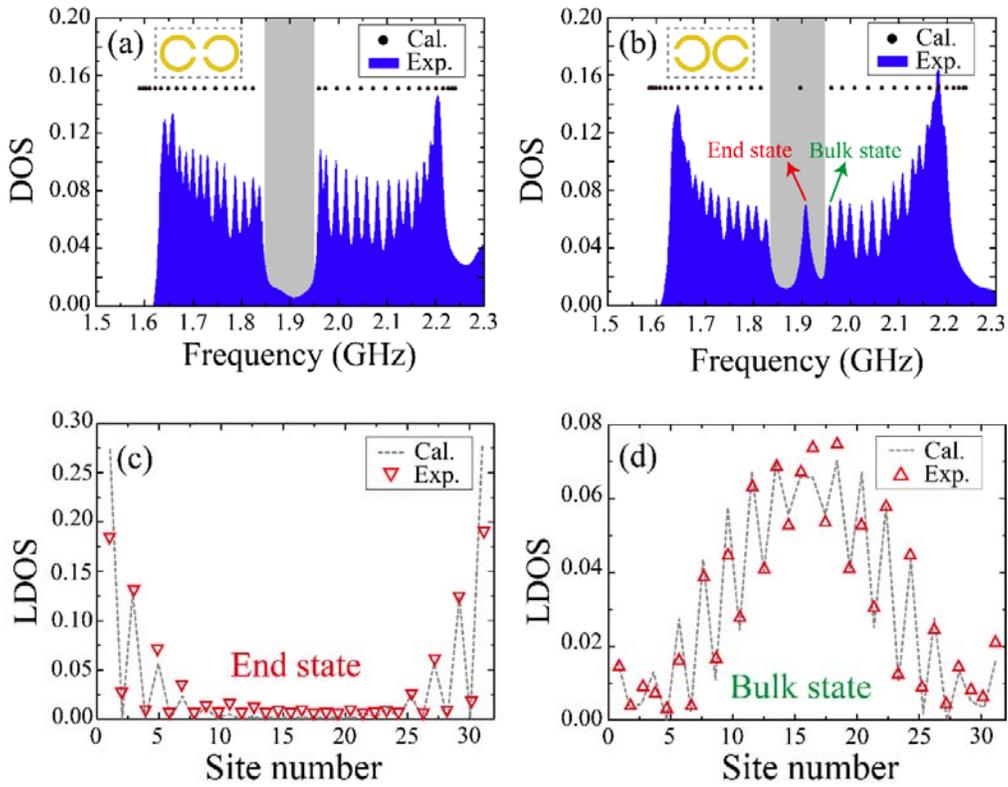

Fig. 2

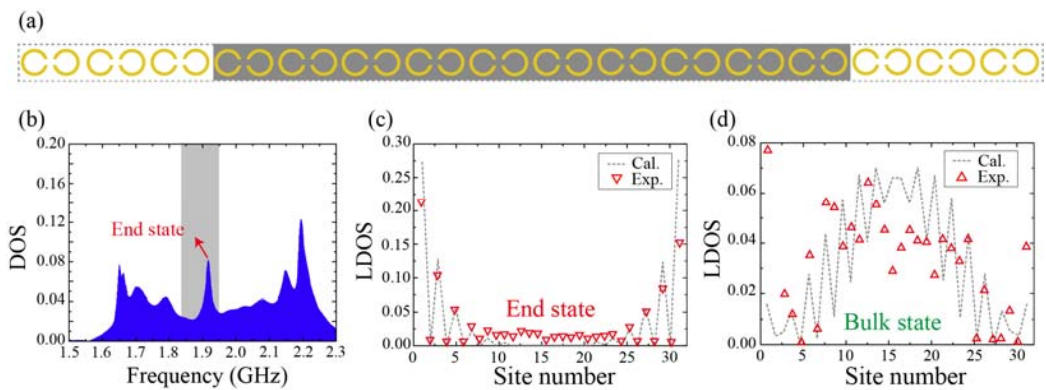

Fig. 3

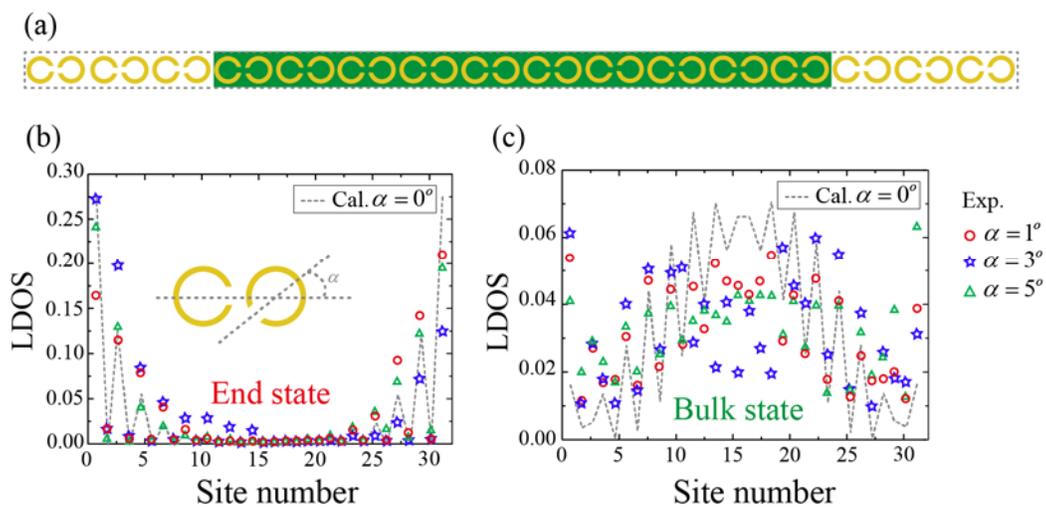

Fig. 4

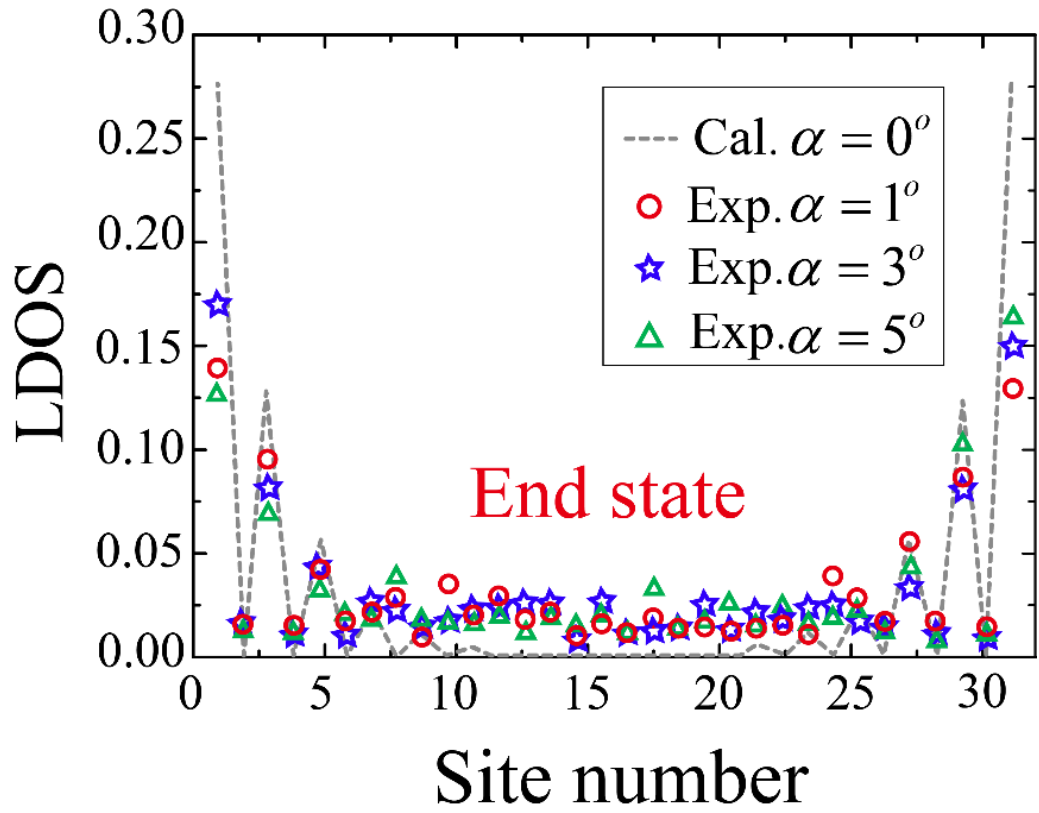

Fig. 5